\documentclass[english]{article}
\usepackage[T1]{fontenc}
\usepackage[latin9]{inputenc}
\usepackage{geometry}
\usepackage{amsmath}
\usepackage{amssymb}
\usepackage{dsfont}
\usepackage{mathtools}
\usepackage{xcolor}
\usepackage{cite}
\usepackage[hidelinks]{hyperref}
\usepackage{physics}
\usepackage{tikz}
\usepackage{slashed}
\usepackage{bbm}

\newcommand{\sa}{\mathsf{a}}

\numberwithin{equation}{section}

\makeatletter
\renewcommand\[{\begin{equation}}
\renewcommand\]{\end{equation}}

\let\mc=\mathcal

\makeatother

\usepackage{babel}
\title{\bfseries{The Schur Expansion of Characteristic Polynomials \\ and Random Matrices}}
\author{\textsc{Taro Kimura}%
\footnote{\texttt{taro.kimura@u-bourgogne.fr}} \space \& \textsc{Edward A. Mazenc}%
\footnote{\texttt{mazenc@uchicago.edu}} }
\date{}

\begin{document}

\maketitle

\vspace{0.5cm}

\begin{center}
   \it{ * Institut de Math\'{e}matiques de Bourgogne,
Universit\'{e} Bourgogne Franche-Comt\'{e}, 
France}\\[.5em]
    \it{$\dagger$ Kadanoff Center for Theoretical Physics, University of Chicago, Chicago, IL 60637 } 
\end{center}

\vspace{2cm}

\begin{abstract}
\noindent We develop a new framework to compute the exact correlators of characteristic polynomials, and their inverses, in random matrix theory. Our results hold for general potentials and incorporate the effects of an external source. In matrix model realizations of string theory, these correspond to correlation functions of exponentiated ``(anti-)branes'' in a given background of ``momentum branes''. Our method relies on expanding the (inverse) determinants in terms of Schur polynomials, then re-summing their expectation values over the allowed representations of the symmetric group. Beyond unifying previous, seemingly disparate calculations, this powerful technique immediately delivers two new results: 1) the full finite $N$ answer for the correlator of inverse determinant insertions in the presence of a matrix source, and 2) access to an interesting, novel regime $M>N$, where the number of inverse determinant insertions $M$ exceeds the size of the matrix $N$. 
\end{abstract}

\newpage

\tableofcontents
\vspace{1em}
\hrule

\section{Introduction and Summary of Results}

Characteristic polynomials, and their inverses, serve as detailed probes of systems described by random matrix ensembles. Indeed, while expectation values of simple traces neatly encapsulate collective properties of the eigenvalues, determinant insertions can capture single eigenvalue effects. For example, their one point functions encode the effective potential for one-eigenvalue instantons \cite{Aganagic_2005,Seiberg_2004}.

The finite $N$ correlation functions of characteristics polynomials have been known for quite some time. They can be written in terms of a determinant built out of the relevant orthogonal polynomials $P_{k}(x)$~\cite{Morozov:1994hh,Brezin:2000CMP},
\begin{equation} \label{eq:justdets}
    \frac{1}{\mc{Z}_{N}} \int \dd{X} e^{- \Tr V(X) } \prod_{\alpha=1}^{M} \det(z_{\alpha} -X) = \frac{\displaystyle \det_{1\le\alpha,\beta\le M}\left[ P_{\alpha + N -1}(z_{\beta}) \right] }{\Delta_{M}(Z)}
\end{equation}
where $X$ is a size $N$ Hermitian matrix, $P_k(x)$ is a monic orthogonal polynomial of degree $k$~\eqref{eq:ortho_polynom}, and
\begin{align}
    \Delta_{M}(z) 
    = \prod_{\alpha<\beta}^M(z_{\alpha}-z_{\beta})
    = \det_{1 \le \alpha, \beta \le M} \qty[ p_{M-\alpha}(z_\beta) ]
    \label{eq:VanderMonde_det}
\end{align}
is the Van-der-Monde determinant for the matrix $Z = \text{diag}(z_{1},...,z_{M})$,%
\footnote{Not to be confused with the partition function $\displaystyle \mc{Z}_{N} = \int \dd{X} e^{- \Tr V(X) } $. See Eq.~\eqref{eq:Normalization_ZN} for the normalization of the integral applied in this paper.} with $p_k(z) = z^k + \cdots$ an arbitrary monic polynomial of degree $k$.

From the string theory perspective, the addition of determinants inside the matrix integral corresponds to adding coherent states of branes, via the immediate relation to the loop operator $\det(z-X)=\exp\left( \Tr \log(z-X) \right)$. Taking the inverse determinant corresponds simply to changing the sign in the exponent; in other words to negative tension branes, sometimes dubbed ``anti-branes''.\footnote{Another reason for this name comes from the fact that \begin{equation}
    \lim_{y \rightarrow z} \frac{\det(z-X)}{\det(y-X)} = 1
\end{equation}
The two insertions thus cancel each other out as we take a coincident limit, much the same way that a particle and its anti-particle annihilate each other.} 
A similar determinantal formula composed out of the Hilbert transform of said orthogonal polynomials computes the correlator of inverse characteristic polynomials~\cite{Fyodorov:2002jw,Strahov:2002zu,Baik:2003JMP,Borodin:2006CPAM}:
\begin{equation} \label{eq:justinversedets}
    \frac{1}{\mc{Z}_{N}} \int \dd{X} e^{- \Tr V(X) } \prod_{\alpha=1}^{M} \det(z_{\alpha} -X)^{-1} = \frac{\tilde{\mc{Z}}_{N-M}}{\tilde{\mc{Z}}_{N}} \frac{\displaystyle \det_{1\le\alpha,\beta\le M} \left[ \tilde{P}_{N-\alpha}(z_{\beta})\right] }{\Delta_{M}(Z)}
\end{equation}
with
\begin{equation}
   \tilde{P}_{k}(z) = \int \frac{\dd{x}}{2\pi} e^{-V(x)} \frac{P_{k}(x)}{z-x} .
   \label{eq:Hilbert_transf}
\end{equation}
See Eq.\eqref{eq:Normalization_ZN} for the definition of the prefactor appearing in Eq.~\eqref{eq:justinversedets}. 

Since the orthogonal polynomials cannot carry negative powers, the $\alpha$ dependence on the RHS of \eqref{eq:justinversedets} immediately begs the question as to what happens when $M>N$. We will see this known formula indeed fails there. As far as we are aware, the case of $M>N$ has been overlooked. Our method allows us to probe this novel regime, giving instead one of the main technical results of this paper
\begin{align}
    \expval{\prod_{\alpha=1}^{M} \det(z_{\alpha}-X)^{-1}} 
    = \frac{\tilde{\mathcal{Z}}_N^{-1}}{\Delta_M(Z)}
    \det_{\substack{i = 1,\ldots,N \\ \alpha = 1,\ldots,M \\ a = 1,\ldots,M-N}}
  \begin{bmatrix}
    \tilde{p}_{N-i}(z_\alpha) \\ p_{a-1}(z_\alpha)
  \end{bmatrix} \qquad (M \geq N). \label{eq:invdetsnosource_M>N}
\end{align}
with the Hilbert transform
\begin{equation}
   \tilde{p}_{k}(z) = \int \frac{\dd{x}}{2\pi} e^{-V(x)} \frac{p_{k}(x)}{z-x} .
\end{equation}

Adding an external source to a matrix ensemble provides a powerful new tool in studying random matrix theory (RMT)~\cite{Brezin:2016eax}. First, it serves as a generating function for correlation functions, as familiar from quantum field theory. Secondly, taking a Wignerian perspective and treating the random matrix as some quantum mechanical Hamiltonian, it allows us to study the effects of random perturbations around some fixed known Hamiltonian, the external source in the matrix integral. The applications do not stop there.
For example, by adding a Gaussian potential centered about the source, Ref.~\cite{blommaert2021gravity} has also recently studied the transition between RMT and a fixed Hamiltonian in the context of Jackiw-Teitelboim (JT) gravity. Sending the width of the Gaussian to zero effectively collapses the matrix integral of \cite{saad2019jt}, picking out the source as the Hamiltonian. In yet another guise, Refs.~\cite{Dijkgraaf_2002,Aganagic_2005} pioneered the use of an external source in the topological string context. They viewed it as a way of inserting branes, who boundary parameter was related to that of the usual determinantal brane via a Fourier transform. These are sometimes referred to as ``momentum branes''
\begin{equation} \label{eq:justsource}
    \int \dd{X} e^{- \Tr V(X) + i \Tr A X} = \frac{C_N}{\Delta_{N}(A)} \det_{1 \le i,j \le N} \left[ Q_{N-i}(\sa_{j})\right] 
\end{equation}
with the constant $C_N$ defined in \eqref{eq:CN_const} and the transformed orthogonal polynomial $Q_{k}(\sa)$ defined as
\begin{equation} \label{eq:legendretransf}
    Q_{k}(\sa)= \int \frac{\dd{x}}{2\pi} e^{-V(x)+i \sa x}p_{k}(x) .
\end{equation}
See, for example, \cite{Morozov:1994hh} for more technical details.
This Fourier transform in Eq.~\eqref{eq:legendretransf} shows in what sense Eq.~\eqref{eq:justsource} represents the insertion of ``momentum branes'' if we interpret Eq.~\eqref{eq:justdets} as ``position brane'' correlators.

It is thus natural to consider an ensemble with both an external source and the addition of (inverse) determinants. In Ref.~\cite{Kimura_2014}, the characteristic polynomial average in the presence of an external matrix source was computed, generalizing the formulas above to
\begin{align}
    \int \dd{X} e^{- \Tr V(X) + i \Tr A X}\prod_{\alpha=1}^{M} \det(z_{\alpha} -X) = & \frac{C_{N}}{\Delta_M(Z) \Delta_{N}(A)} \det_{\substack{1 \le \alpha,\beta \le M \\ 1 \le i, j \le N}}
    \begin{bmatrix}
    p_{N + M - \beta}(z_\alpha) & p_{N - j}(z_{\alpha}) \\
    Q_{N+M-\beta}(\sa_{i}) & Q_{N - j}(\sa_{i})
    \end{bmatrix} .
    \label{eq:ChPoly_av_A}
\end{align}
See Section~\ref{sec:detwithsource} for the details.

In this paper, we will reproduce all the above formula using the Schur expansion of the (inverse) characteristic polynomial. Our method therefore unifies the otherwise seemingly disparate techniques used to obtain the results above. Furthermore, it will allow us to compute the inverse determinant correlators in the presence of a source, first for the case $M \leq N$:
\begin{align}
   \int \dd{X} e^{- \Tr V(X) + i \Tr A X}\prod_{\alpha=1}^{M} \det(z_{\alpha} -X)^{-1}  = & \frac{C_{N}}{\Delta_M(Z)\Delta_N(A)} \det_{\substack{1\le\alpha,\beta\le M \\ M+1\le a,b \le N}} 
    \begin{bmatrix}
    \tilde{Q}(\sa_\alpha,z_\beta) & Q_{N-b}(\sa_\alpha) \\ \tilde{Q}(\sa_a,z_\beta) & Q_{N-b}(\sa_a)
    \end{bmatrix} 
    \label{eq:inv_ch_poly_A}
\end{align}
with the new function $\tilde{Q}(\sa,z)$ defined as
\begin{equation} \label{eq:tildeQ}
    \tilde{Q}(\sa,z) = \int \frac{\dd{x}}{2\pi} e^{- V(x)+i \sa x} \frac{1}{z-x} .
\end{equation}
while for $M \geq N$, we obtain a generalization of Eq.~\eqref{eq:invdetsnosource_M>N},
\begin{align}
    \expval{\prod_{\alpha=1}^{M} \det(z_{\alpha}-X)^{-1}}_{A} 
    = \frac{C_N}{\Delta_M(Z) \Delta_N(A)} \det_{\substack{i = 1,\ldots,N \\ \alpha = 1,\ldots,M \\ a = 1,\ldots,M-N}}
  \begin{bmatrix}
    \tilde{Q}(\sa_i,z_\alpha) \\ p_{a-1}(z_\alpha)
  \end{bmatrix}  \qquad (M \geq N).
\end{align}
Together, these two formula constitute the other main new result of this work, and a useful application of our Schur polynomial method.

The goal of this paper is to provide a usable introduction to the application of Schur polynomials towards computing exact determinant and inverse determinant correlators in RMT. We have chosen a rather pedagogical tone, spelling out several of the intermediate steps, as they may not be familiar to many readers. 

We begin with the study of characteristic polynomials, and their inverses, without external matrix source. Section \ref{sec:schurpolyaverage} introduces the Schur polynomial expansion and computes its expectation value.  In sections \ref{sec:detnosource} and \ref{sec:invnosource}, we perform the sum over the relevant representations of the symmetric group, reproducing the familiar formulas discussed above. Section \ref{sec:withsource} generalizes this approach to non-zero source. In particular, section \ref{sec:invwithsource} finds the exact correlator of inverse determinant insertions in this ensemble, a computation which had resisted previous methods. We conclude with a discussion focused around the interpretation of the $M>N$ result and a string theory perspective on the Schur expansion.

\section{Characteristic Polynomials without Source} \label{sec:nosource} 

\subsection{Schur Polynomial Average} \label{sec:schurpolyaverage}

We first explore the Schur polynomial method in the more familiar context of RMT without an external field.
The Schur polynomial $s_{\lambda}(X)$, parametrized by a partition $\lambda = (\lambda_1 \ge \lambda_2 \ge \cdots \ge \lambda_N) \in \mathbb{Z}_{\ge 0}^N$, has a determinantal expression 
\begin{equation}
    s_{\lambda}(X) = s_\lambda(x_1,\ldots,x_N) =\frac{\displaystyle \det_{1\le i,j\le N}\left[ x_{i}^{\lambda_{j}+N-j}\right]}{\Delta_N(X)} .
\end{equation}
We will need to evaluate the Schur polynomial average,
\begin{align}
    \expval{s_\lambda(X)}
    & = \frac{1}{\mc{Z}_N} \int \dd{X} e^{-\Tr V(X)} s_\lambda(X) .
\end{align}
Diagonalizing the Hermitian matrix $X$, and applying expression~\eqref{eq:VanderMonde_det}, we obtain%
\footnote{
Diagonalization of the rank $N$ Hermitian matrix measure gives rise to
\begin{align}
    \int \dd{X} = \frac{\operatorname{Vol} U(N)}{|\mathfrak{S}_N| \times \operatorname{Vol} U(1)^N} \int \prod_{i=1}^N \dd{x}_i \Delta_N(X)^2 ,
\end{align}
where $\mathfrak{S}_N$ is the degree $N$ symmetric group to be identified with the Weyl group of $U(N)$, and $U(1)^N$ is the maximal Cartan torus of $U(N)$ with $|\mathfrak{S}_N| = N!$ and $\operatorname{Vol} U(1) = 2\pi$.
See~\cite{Eynard:2015aea} for details.
}
\begin{align}
    \expval{s_\lambda(X)} & = \frac{\operatorname{Vol} U(N)}{N! \mc{Z}_N} \int \prod_{i=1}^N \frac{\dd{x}_i}{2\pi} e^{-V(x_i)} \det_{1 \le i, j \le N} \left[ p_{N-j}(x_i) \right] \det_{1 \le i,j \le N} \left[ x_i^{\lambda_j + N - j}\right]
    ,
\end{align}
where the volume of $U(N)$ group is given by the double gamma function,
\begin{align}
    \operatorname{Vol} U(N) = \frac{(2 \pi)^{N(N+1)/2}}{\Gamma_2(N+1)}
    \, , \qquad
    \Gamma_2(N+1) = \prod_{k=1}^{N-1} k!
    \, .
\end{align}
Now we may apply the Andr\'{e}if--Heine identity
\begin{align}
    \frac{1}{N!} \int \prod_{i=1}^N \dd{x}_i \det_{1 \le i, j \le N} \left[ \psi_{j} (x_i) \right]  \det_{1 \le i, j \le N} \left[ \phi_{j} (x_i) \right] 
    =  \det_{1 \le i, j \le N} \qty[ \int \dd{x} \psi_i(x) \phi_j(x) ] ,
    \label{eq:AH_id}
\end{align}
which gives rise to the determinantal formula for the Schur polynomial average,
\begin{align}
    \expval{s_\lambda(X)} = \frac{\operatorname{Vol} U(N)}{\mc{Z}_N} \det_{1 \le i, j \le N} \qty[\braket{p_{N - i}}{x^{\lambda_j + N - j}}]
\end{align}
where we define the inner product together with the weight function $e^{-V(x)}$,
\begin{align}
    \braket{\psi}{\phi} = \int \frac{\dd{x}}{2\pi} \, e^{-V(x)} \psi(x) \phi(x)
    \, .
\end{align}

\subsubsection*{Normalization}

For the trivial partition $\lambda = \emptyset$, we obtain
\begin{align}
    \expval{s_\emptyset(X)} 
    = \frac{\operatorname{Vol} U(N)}{\mc{Z}_N} \det_{1 \le i, j \le N} \qty[\braket{p_{N - i}}{x^{N - j}}]
    = \frac{\operatorname{Vol} U(N)}{\mc{Z}_N} \det_{1 \le i, j \le N} \qty[\braket{ p_{N - i} }{ p_{N-j} }]
    \, .
\end{align}
Taking the orthogonal polynomial as a monic polynomial $p_k(x) \to P_k(x)$, such that
\begin{align}
    \braket{P_i}{P_j} = h_i \, \delta_{i,j}
    \, ,
    \label{eq:ortho_polynom}
\end{align}
the Schur polynomial average reads
\begin{align}
    \expval{s_\emptyset(X)} 
    = \frac{\operatorname{Vol} U(N)}{\mc{Z}_N} \prod_{i=1}^N h_{i-1} .
\end{align}
Recalling $s_\emptyset(X) = 1$, we obtain the partition function
\begin{align}   \label{eq:Normalization_ZN}
    \mc{Z}_N = \operatorname{Vol} U(N) \tilde{\mc{Z}}_N
    \, , \qquad
    \tilde{\mc{Z}}_N = \prod_{i=1}^N h_{i-1} ,
\end{align}
so that
\begin{align}
    \expval{s_\lambda(X)} = \frac{1}{\tilde{\mc{Z}}_N} \det_{1 \le i, j \le N} \qty[\braket{p_{N - i}}{x^{\lambda_j + N - j}}] .
    \label{eq:SchurPoly_av}
\end{align}

\subsection{Characteristic Polynomial Average} \label{sec:detnosource}

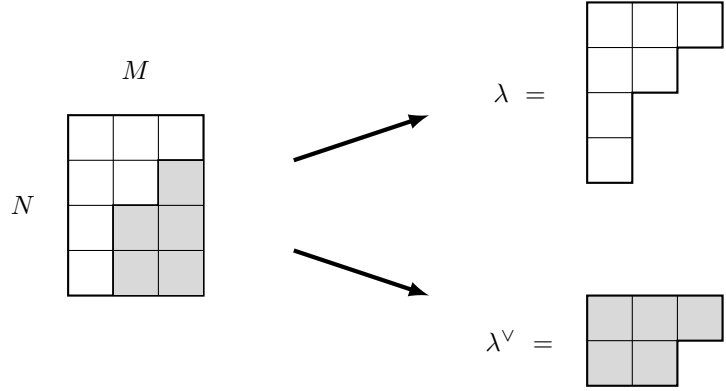
\begin{figure}[ht]
\centering
    \begin{tikzpicture}[scale=.6,baseline=(current  bounding box.center)]
  \draw[thick] (0,0) rectangle ++(3,-4);
  \draw[thick] (3,-1) -- ++(-1,0) -- ++(0,-1) -- ++(-1,0) -- ++(0,-2);
  \filldraw [fill=gray!30] (3,-1) -- ++(-1,0) -- ++(0,-1) -- ++(-1,0) -- ++(0,-2) -- ++(2,0) -- cycle;
  \draw (0,-1) -- ++(3,0);
  \draw (0,-2) -- ++(3,0);
  \draw (0,-3) -- ++(3,0);
  \draw (1,0) -- ++(0,-4);
  \draw (2,0) -- ++(0,-4);    
  \node at (-1,-2) {$N$};
  \node at (1.5,1) {$M$};  
  \draw [ultra thick,-latex] (5,-1) -- ++(3,1);
  \draw [ultra thick,-latex] (5,-3) -- ++(3,-1);
   \begin{scope}[shift={(11.5,2.5)}]
    \node at (-1.5,-2) {$\lambda \ =$};
    \draw [thick] (0,0) -- ++(3,0) -- ++(0,-1) -- ++(-1,0) -- ++(0,-1) -- ++(-1,0) -- ++(0,-2) -- ++(-1,0) -- cycle;
  \draw (0,-1) -- ++(3,0);
  \draw (0,-2) -- ++(2,0);
  \draw (0,-3) -- ++(1,0);
  \draw (1,0) -- ++(0,-2);
  \draw (2,0) -- ++(0,-1);    
   \end{scope}
  \begin{scope}[shift={(11.5,-4.)}]
   \node at (-1.5,-1) {$\lambda^\vee \ =$};
   \filldraw [thick,fill=gray!30] (0,0) -- ++(3,0) -- ++(0,-1) -- ++(-1,0) -- ++(0,-1) -- ++(-2,0) -- cycle;
   \draw (0,-1) -- ++(2,0);
   \draw (1,0) -- ++(0,-2);
   \draw (2,0) -- ++(0,-2);
   \draw (3,0) -- ++(0,-1);      
  \end{scope}
\end{tikzpicture}
\caption{An example of a partition $\lambda$ and its dual $\lambda^\vee$ for the case of $N=4$ and $M=3$: $\lambda=(3,2,1,1)$ and $\lambda^\vee=(3,2,0)$, while $\lambda^\text{T}=(4,2,1)$.} \label{fig:dualpart}
\end{figure}

Based on the Schur polynomial average discussed above, we wish to compute the following correlation functions exactly in $N$ and $M$,
\begin{equation} \label{eq:goalnosource_chpoly}
    \expval{\prod_{\alpha=1}^{M} \det(z_{\alpha}-X)}=\frac{1}{\mc{Z}_{N}}\int \dd{X} e^{-\Tr V(X)} \prod_{\alpha=1}^{M} \det(z_{\alpha}-X) .
\end{equation}
For this purpose, we apply the Schur polynomial expansion of the characteristic polynomial product
\begin{align} \label{eq:Schurexpdet}
    \prod_{\alpha=1}^M \det(z_\alpha - X) = \sum_{\lambda \subseteq (M^N)} (-1)^{|\lambda|} s_{\lambda^\vee} (Z) s_\lambda(X)
\end{align}
where $Z = (z_1,\ldots,z_M)$ and $X = (x_1,\ldots,x_N)$.
This expansion is due to the Cauchy sum formula of the Schur polynomial~\cite{Macdonald:1997}.
We denote the transposition of the partition $\lambda$ by $\lambda^\text{T}$.
Then, the dual partition $\lambda^\vee$ is defined as follows:
\begin{align}
    \lambda^\vee = (\lambda_1^\vee,\lambda_2^\vee,\ldots,\lambda_M^\vee) = (N - \lambda_M^\text{T}, N - \lambda_{M-1}^\text{T}, \ldots, N - \lambda_1^\text{T})
    \, ,
\end{align}
which is graphically expressed in Fig.~\ref{fig:dualpart}.
Using the Schur polynomial average~\eqref{eq:SchurPoly_av}, the characteristic polynomial correlator is given by
\begin{align}
    \expval{\prod_{\alpha=1}^N \det(z_\alpha - X)}
    & = \frac{1}{\Delta_M(Z)} \sum_{\lambda \subseteq (M^N)} (-1)^{|\lambda|} \det_{1 \le \alpha,\beta \le M} \qty[ z_\alpha^{\lambda^{\vee}_{\beta}+M-\beta} ] \expval{s_\lambda(X)}
    \nonumber \\
    & = \frac{\tilde{\mathcal{Z}}_N^{-1} }{\Delta_M(Z)} \sum_{\lambda \subseteq (M^N)} (-1)^{|\lambda|} \det_{1 \le \alpha,\beta \le M} \qty[ z_\alpha^{\lambda^{\vee}_{\beta}+M-\beta} ] \det_{1 \le i, j \le N} \qty[ \braket{p_{N - i}}{x^{\lambda_j + N - j}} ] .
\end{align}

In order to carry out the summation over partitions $\lambda$, we first briefly recall the generalized co-factor expansion of an $(N+M)\times(N+M)$ matrix $B$ along the first $M$ rows. Let $S$ by the set of all $M$-element subsets of $\{1,2,...,N+M\}$, $H$ the set $\{1,..,M\}$ (this is our choice of expanding along the first $M$ rows) and $H'$ the complement of $H$ (the set $\{M+1,..,N+M\}$). The expansion then reads
\begin{equation} \label{eq:cofactorexp}
   \det(B) = \sum_{L \in S} (-1)^{(\sum_{h \in H} h + \sum_{l \in L}l)} \det(B_{H,L}) \det(B_{H',L'}) 
\end{equation}
where $B_{H,L}$ is the square minor obtained by deleting rows and columns in $H$ and $L$ respectively. $B_{H',L'}$ is similarly defined, with $L'$ the complement of $L$, i.e. $L \cup L' = S$. 

We now wish to identify the summation over the partition $\lambda$ as a rank $M$ co-factor expansion of the following rank $(N+M)$ determinant,
\begin{align} \label{eq:N+Mblockjustdets}
    &
    \sum_{\lambda \subseteq (M^N)} (-1)^{|\lambda|} \det_{1 \le \alpha,\beta \le M} \qty[ z_\alpha^{\lambda^{\vee}_{\beta}+M-\beta} ] \det_{1 \le i, j \le N} \qty[ \braket{p_{N - i}}{x^{\lambda_j + N - j}} ]
    \nonumber \\
    & = \det_{\substack{1 \le \alpha,\beta \le M \\ 1 \le i, j \le N}}
    \begin{bmatrix}
    z_\alpha^{N + M - \beta} & z_{\alpha}^{N - j} \\
    \braket{p_{N - i}}{x^{N + M - \beta}} & \braket{p_{N - i}}{x^{N - j}}
    \end{bmatrix}
    = \det_{\substack{1 \le \alpha,\beta \le M \\ 1 \le i, j \le N}}
    \begin{bmatrix}
    p_{N + M - \beta}(z_\alpha) & p_{N - j}(z_{\alpha}) \\
    \braket{p_{N - i}}{p_{N + M - \beta}} & \braket{p_{N - i}}{p_{N - j}}
    \end{bmatrix}
    \, .
\end{align}
where in going to the last equality, we used the linearity of the inner-product and the invariance of the determinant under linear transformations on its rows to trade $x^{N+M-\beta} \rightarrow p_{N+M-\beta}(x)$, and similarly for $z$.

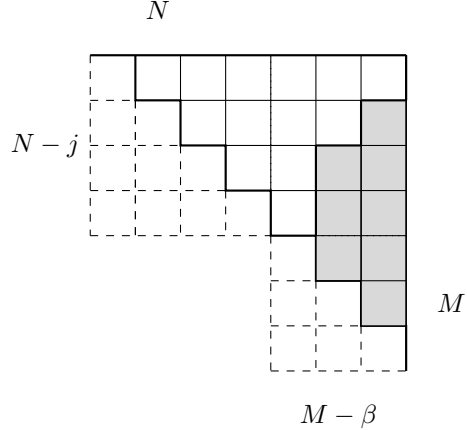
\begin{figure}[ht]
\centering
    \begin{tikzpicture}[scale=.6,baseline=(current  bounding box.center)]
  \path[dashed] (0,0) rectangle ++(7,-7);
  \draw[thick] (7,-7) -- ++ (0,1) -- ++(-1,0) -- ++ (0,1) -- ++ (-1,0) -- ++ (0,1) -- ++ (-1,0) -- ++ (0,1) -- ++ (-1,0) -- ++ (0,1) -- ++ (-1,0) -- ++ (0,1) -- ++ (-1,0) -- ++ (0,1);
  \filldraw [fill=gray!30] (7,-1) -- ++(0,1) -- ++(0,-6) -- ++(-1,0) -- ++(0,5);
  \filldraw [fill=gray!30] (6,-2) -- ++(0,1) -- ++(0,-4) -- ++(-1,0) -- ++(0,3);
  \draw[dashed] (0,-1) -- ++(1,0);
  \draw[dashed] (0,-2) -- ++(2,0);
  \draw[dashed] (0,-3) -- ++(3,0);
  \draw[dashed] (0,-4) -- ++(4,0);
  \draw[dashed] (4,-4) -- ++(0,-3);
  \draw[dashed] (5,-5) -- ++(0,-2);
  \draw[dashed] (6,-6) -- ++(0,-1);
  \draw[dashed] (4,-5) -- ++(1,0);
  \draw[dashed] (4,-6) -- ++(2,-0);
  \draw[dashed] (4,-7) -- ++(3,-0);
  \draw[dashed] (0,0) -- ++(0,-4);
  \draw[dashed] (1,-1) -- ++(0,-3);
  \draw[dashed] (1,-1) -- ++(0,-3);
  \draw[dashed] (2,-2) -- ++(0,-2);
  \draw[dashed] (3,-3) -- ++(0,-1);
  \draw (7,0) -- ++(0,-7);
  \draw (6,0) -- ++(0,-6);
  \draw (5,0) -- ++(0,-5);
  \draw (4,0) -- ++(0,-4);
  \draw (3,0) -- ++(0,-3);
  \draw (2,0) -- ++(0,-2);
  \draw (0,0) -- ++(7,0);
  \draw (7,0) -- ++(0,-7);
  \draw (1,-1) -- ++(6,0);
  \draw (2,-2) -- ++(5,0);
  \draw (3,-3) -- ++(4,0);
  \draw (4,-4) -- ++(3,0);
  \draw (5,-5) -- ++(2,0);
  \draw[thick] (7,0) -- ++(0,-1);
  \draw[thick] (7,-1) -- ++(-1,0);
  \draw[thick] (6,-1) -- ++(0,-1);
  \draw[thick] (6,-2) -- ++(-1,0);
  \draw[thick] (5,-2) -- ++(0,-2);
  \draw[thick] (0,0) -- ++(7,0);
  \draw[dotted] (4,0) -- ++(0,-3);
  \draw[dotted] (5,-4) -- ++(2,0);
  \node at (-1,-2) {$N-j$};
  \node at (1.5,1) {$N$};
  \node at (8,-5.5) {$M$};
  \node at (5.5,-8) {$M-\beta$};
\end{tikzpicture}  
\caption{We can append to the Young Tableau of Fig.~\ref{fig:dualpart} an $N\times N$ square block to left and an $M \times M$ block below. The number of total white squares in each row are labeled by $\lambda_{j}+N-j$. The number of total shaded squares in each column are labeled by $\lambda^{\vee}_{\beta}+M-\beta$. For this particular example of partition $\lambda$, we would thus have $B_{H',L'} = \braket{p_{N-i}}{x^{a}}$ for $a \in \{1,2,4,6\}$ and $B_{H,L}= z_{\beta}^{c}$ for $c \in \{0,3,5\}$.} \label{fig:extendedYT}
\end{figure}

Remark that $\lambda \subseteq (M^N)$, so that $0 \le \lambda_j + N - j \le N + M - 1$ for $j = 1,\ldots,N$. In other words, there are always $\lambda_{j}$ such that we may obtain any $N$-column subset of the matrix $\braket{p_{N - i}}{x^{N+M -a}}$. The sum over $\lambda$ is thus indeed a sum over the different subsets of $N$ columns of the matrix $\braket{p_{N - i}}{x^{N+M -a}}$. We may rewrite the rank-$N$ matrix as
\begin{align}
    \qty[\braket{p_{N - i}}{x^{\lambda_j + N - j}}]_{i,j = 1,\ldots,N}
    = \begin{bmatrix} \bra{p_{N-1}} \\ \vdots \\ \bra{p_0} \end{bmatrix} 
    \qty[ \ket{x^{N + M - 1}} \cdots 
    | \widehat{x^{\lambda_1^\vee + M - 1}} \rangle
    \cdots 
    | \widehat{x^{\lambda_M^\vee}} \rangle
    \cdots \ket{x^0} ]
\end{align}
where the hat is a removal symbol. This reflects the fact that $\lambda^\vee$ plays the same role as $L$ in \eqref{eq:cofactorexp} (the elements $l \in L$ are related via $l \leftrightarrow N+\beta - \lambda^{\vee}_{\beta}$). Finally, we used that $NM-|\lambda^{\vee}|=|\lambda|$ to match the grading appearing in Eq. \ref{eq:cofactorexp}. We may express all this graphically as in Fig.~\ref{fig:extendedYT}. 

Taking a monic polynomial as the orthogonal polynomial~\eqref{eq:ortho_polynom} as before, the lower left block in Eq.~\eqref{eq:N+Mblockjustdets} vanishes and the determinant factorizes into the product of determinants for each block. We thus obtain the determinantal formula of the characteristic polynomial average,
\begin{align}
    \expval{\prod_{\alpha=1}^N \det(z_\alpha - X)}
    & = 
    \frac{1}{\Delta_M(Z)} \det_{1 \le \alpha,\beta \le M} \qty[ P_{N + \alpha - 1}(z_\beta) ]
    \, ,
\end{align}
which reproduces formula~\eqref{eq:justdets}.

\subsection{Inverse Characteristic Polynomial Average} \label{sec:invnosource}

We now wish to compute the following correlation functions, exactly in $N$ and $M$,
\begin{equation} \label{eq:goalnosource}
    \expval{\prod_{\alpha=1}^{M} \det(z_{\alpha}-X)^{-1}}=\frac{1}{\mc{Z}_{N}}\int \dd{X} e^{-\Tr V(X)} \prod_{\alpha=1}^{M} \frac{1}{\det(z_{\alpha}-X)} .
\end{equation}
The Cauchy sum formula also allows us to expand the inverse characteristic polynomial in terms of a sum of products of Schur polynomials
\begin{equation} \label{eq:CauchyIdentity}
    \prod_{\alpha=1}^{M} \frac{1}{\det(z_{\alpha}-X)} = \frac{1}{\det (Z)^N} \sum_{\ell(\lambda)\leq \min(M,N)} s_{\lambda}(Z^{-1})s_{\lambda}(X)
\end{equation}
where $X=(x_{1},..,x_{N})$ are the eigenvalues of the matrix $X$ and $\ell(\lambda) = \lambda^\text{T}_1$. This is a sum over a subset of representations of the symmetric group. 
We will first assume $M<N$, so that $\lambda_{i}=0$ for all $i>M$, i.e. the relevant Young tableaux all have at most $M$ rows, $\ell(\lambda)\leq M$.

Note we can rewrite part of Eq.~\eqref{eq:CauchyIdentity} as 
\begin{equation}
    \frac{s_{\lambda}(Z^{-1})}{\det(Z)^{N}} = \frac{\displaystyle \det_{1\le \alpha,\beta \le M} \left[ z_{\alpha}^{-\lambda_{\beta}+\beta-(N+1)} \right] }{\Delta_M(Z)} .
    \label{eq:Schur_inv}
\end{equation}
While we have already computed the expectation value of $s_{\lambda}(x)$, 
we now want to exploit the fact that $\lambda_{i}=0$ for $i>M$. To make this split manifest, let Greek letters $\alpha, \beta... \in (1,\ldots,M)$ and reserve early Latin alphabet letters $a,b \in (M+1,\ldots,N)$. Then we may rewrite Eq.~\eqref{eq:SchurPoly_av} in block form 

\begin{align}
     \expval{s_\lambda(X)} 
    & = 
    \frac{1}{\tilde{\mc{Z}}_{N}} \det_{\substack{1 \le \alpha, \beta \le M \\ M+1 \le a,b \le N}}
    \begin{bmatrix}
    \braket{p_{N-\alpha}}{x^{\lambda_\beta + N - \beta}} & \braket{p_{N-\alpha}}{p_{N - b}} \\
    \braket{p_{N-a}}{x^{\lambda_\beta + N - \beta}} & \braket{p_{N-a}}{p_{N - b}}
    \end{bmatrix}
    \nonumber \\ & =
    \frac{\tilde{\mc{Z}}_{N-M}}{\tilde{\mc{Z}}_{N}}
    \det_{1 \le \alpha, \beta \le M} \qty[ \braket{P_{N-\alpha}}{x^{\lambda_\beta + N - \beta}} ]
    \label{eq:SchurPoly_av_block}
\end{align}
In going to the second line, we again took $p_{k} \rightarrow P_{k}$ via linear transformations on the rows. The upper right block then vanishes due to their orthogonality and the determinant factorizes. The prefactor $\tilde{\mc{Z}}_{N-M} = \prod_{a=M+1}^{N} h_{N-a}$ arises as the determinant of the lower right block.

Having computed the expectation value of the Schur polynomial, we can now write Eq.~\eqref{eq:goalnosource} as 
\begin{align} \label{eq:sumoverreps}
    & \expval{\prod_{\alpha=1}^{M} \det(z_{\alpha}-X)^{-1}} \nonumber \\ & 
    = \frac{ \tilde{\mc{Z}}_{N-M}/\tilde{\mc{Z}}_{N} 
    }{\Delta_M(Z)} \sum_{0\leq \lambda_{M}\leq...\leq \lambda_{1} \leq \infty} \det_{1 \le \alpha, \beta \le M} \left[ z_{\alpha}^{-\lambda_{\beta}+\beta-(N+1)}\right] \det_{1 \le \alpha, \beta \le M} \qty[ \braket{P_{N-\alpha}}{x^{\lambda_\beta + N - \beta}} ] .
\end{align}
To perform this sum, first define the shifted variable $r_{\alpha} = \lambda_{\alpha} + M - \alpha$, and note that even if $\lambda_{\alpha+1}=\lambda_{\alpha}$, $r_{\alpha} > r_{\alpha+1}$, namely $(r_1 > r_2 > \cdots > r_M) \in \mathbb{Z}^M_{\ge 0}$.
We can then rewrite Eq. \eqref{eq:sumoverreps} as  \footnote{The factor of $1/M!$ below comes from the lack of ordering in the sum over $r_{\alpha}$.}
\begin{align}
    \expval{\prod_{\alpha=1}^{M} \det(z_{\alpha}-X)^{-1}} = & \frac{\tilde{\mc{Z}}_{N-M}/\tilde{\mc{Z}}_{N}}{\Delta_M(Z) M!} \sum_{\substack{ 0 \le r_{1},...,r_{M} \le \infty \\ r_{\alpha}\neq r_{\beta}}} \det_{1 \le \alpha, \beta \le M} \left[ z_{\alpha}^{M-N-r_{\beta}-1}\right] \det_{1 \le \alpha, \beta \le M} \qty[ \braket{P_{N - \alpha}}{x^{-M+r_\beta+N}} ]
    \nonumber\\
    = & \frac{\tilde{\mc{Z}}_{N-M}/\tilde{\mc{Z}}_{N}}{\Delta_M(Z)} \det_{1 \le \alpha, \beta \le M} \left[ \sum_{r=0}^{\infty} z_{\alpha}^{M-N-r-1} \braket{P_{N-\beta}}{x^{N - M + r}}
    \right]
\end{align}
where we used a discrete equivalent of the Andr\'{e}if--Heine identity to trade the sum of products of determinants for a determinant of the sum of products. 

We can now explicitly perform the sum over $r$, and recognize it as the geometric series expansion $(z_{\alpha}-x)^{-1}=\sum_{r=0}^{\infty} z^{-r-1}x^{-r}$. This leaves us with 
\begin{equation}
     \expval{\prod_{\alpha=1}^{M} \det(z_{\alpha}-X)^{-1}} = \frac{\tilde{\mc{Z}}_{N-M}/\tilde{\mc{Z}}_{N}}{\Delta_M(Z)} \det \left[ z_{\alpha}^{M-N} \int \frac{\dd{x}}{2\pi} e^{-V(x)}  P_{N-\beta}(x) \frac{x^{N-M} }{z_{\alpha}-x} \right]
\end{equation}
To recover the more familiar expression, first rewrite
\begin{equation}
\frac{x^{N-M} }{z-x} = \frac{z^{N-M}}{z-x}-\frac{z^{N-m}-x^{N-m}}{z-x} = \frac{z^{N-M}}{z-x} + \mathcal{O}(x^{N-M-1})
\end{equation}
and note that, by the orthogonality,
\begin{equation}
    \int \frac{\dd{x}}{2\pi} e^{-V(x)}  P_{N-\beta}(x) \mathcal{O}(x^{N-(M+1)}) = 0 .
\end{equation}
Recalling the definition of the Hilbert transform~\eqref{eq:Hilbert_transf}, this gives us our final expression 
\begin{align}
     \expval{\prod_{\alpha=1}^{M} \det(z_{\alpha}-X)^{-1}} & = \frac{\tilde{\mc{Z}}_{N-M}/\tilde{\mc{Z}}_{N}}{\Delta_M(Z)} \det_{1 \le \alpha, \beta \le M} \left[ \int \frac{\dd{x}}{2\pi} e^{-V(x)}  \frac{ P_{N-\beta}(x) }{z_{\alpha}-x} \right]
     \nonumber \\
     & = \frac{\tilde{\mc{Z}}_{N-M}/\tilde{\mc{Z}}_{N}}{\Delta_M(Z)} \det_{1 \le \alpha, \beta \le M} \left[ \tilde{P}_{N-\beta}(z_\alpha) \right]
\end{align}
which reproduces the classic formula for inverse characteristic polynomials in terms of a determinant over Hilbert-transforms of the orthogonal polynomials, as advertised in Eq.\eqref{eq:justinversedets}.
This shows a different (and also simpler) derivation of the inverse characteristic polynomial formula, originally found in \cite{Fyodorov:2002jw}.

\subsubsection*{The Case with $M>N$}

Let us now consider the opposite situation $M>N$.
In this case, a different manipulation is necessary in the Cauchy sum formula \eqref{eq:CauchyIdentity}. 
We now have the condition $\ell(\lambda) \le \operatorname{min}(N,M) = N$, so that $\lambda_\alpha = 0$ for all $\alpha > N$.

While there are no immediate simplification to be made to our expression for $\expval{s_{\lambda}(X)}$, we may instead exploit the following block matrix structure for the $M$-variable Schur polynomial~\eqref{eq:Schur_inv},
\begin{align}
  \frac{s_{\lambda}(Z^{-1})}{\det(Z)^{N}} = \frac{1}{\Delta_M(Z)} \det_{\substack{i = 1,\ldots,N \\ \alpha = 1,\ldots,M \\ a = 1,\ldots,M-N}}
  \begin{bmatrix}
    z_\alpha^{-\lambda_i + i - (N+1)} \\ z_{\alpha}^{a-1}
  \end{bmatrix} .
  \label{eq:block_Z_M>N}
\end{align}
Thus, the inverse characteristic polynomial average admits the following expansion over partitions,
\begin{align}
    \expval{ \prod_{\alpha=1}^M \det(z_\alpha - X)^{-1} }
    & = \frac{\tilde{\mc{Z}}_N^{-1}}{\Delta_M(Z)} \sum_{0 \le \lambda_N \le \cdots \le \lambda_1 \le \infty} \det_{\substack{i = 1,\ldots,N \\ \alpha = 1,\ldots,M \\ a = 1,\ldots,M-N}}
  \begin{bmatrix}
    z_\alpha^{-\lambda_i + i - (N+1)} \\ z_{\alpha}^{a-1}
  \end{bmatrix} \det_{1 \le i, j \le N} \qty[ \braket{p_{N-i}}{x^{\lambda_j + N - j}} ] .
\end{align}

To perform the summation over the partition, we reintroduce $r_i = \lambda_i + N - i$ obeying the strictly decreasing condition $r_1 > r_2 > \cdots > r_N \ge 0$.
We can use a slight generalization of the discrete Andr\'{e}if--Heine identity accommodating the determinant of different size matrices \footnote{This is easiest to see by expanding the determinants using the Levi-Civita symbol.} to obtain
\begin{align}
    & \sum_{0 \le \lambda_N \le \cdots \le \lambda_1 \le \infty} \det_{\substack{i = 1,\ldots,N \\ \alpha = 1,\ldots,M \\ a = 1,\ldots,M-N}}
  \begin{bmatrix}
    z_\alpha^{-\lambda_i + i - (N+1)} \\ z_{\alpha}^{a-1}
  \end{bmatrix} \det_{1 \le i, j \le N} \qty[ \braket{p_{N-i}}{x^{\lambda_j + N - j}} ]
  \nonumber \\ &
  = \det_{\substack{i = 1,\ldots,N \\ \alpha = 1,\ldots,M \\ a = 1,\ldots,M-N}}
  \begin{bmatrix}
    \displaystyle \sum_{r=0}^\infty z_\alpha^{-r-1} \braket{p_{N-i}}{x^r} \\ z_\alpha^{a-1}
  \end{bmatrix}
  = \det_{\substack{i = 1,\ldots,N \\ \alpha = 1,\ldots,M \\ a = 1,\ldots,M-N}}
  \begin{bmatrix}
    \tilde{p}_{N-i}(z_\alpha) \\ p_{a-1}(z_\alpha)
  \end{bmatrix} ,
\end{align}
where we define the Hilbert transform of the monic polynomial,
\begin{align}
    \tilde{p}_k(z) = \int \frac{\dd{x}}{2\pi} e^{-V(x)} \frac{p_k(x)}{z - x} .
\end{align}
We have also used the invariance of the determinant under linear transformation to convert the monomial $z_\alpha^{a-1}$ to an arbitrary monic polynomial $p_{a-1}(z_\alpha)$.
This gives us the following determinantal formula for the inverse characteristic polynomial in the case of $M > N$,
\begin{align}
    \expval{ \prod_{\alpha=1}^M \det(z_\alpha - X)^{-1} }
    = \frac{\tilde{\mc{Z}}_N^{-1}}{\Delta_M(Z)} 
    \det_{\substack{i = 1,\ldots,N \\ \alpha = 1,\ldots,M \\ a = 1,\ldots,M-N}}
  \begin{bmatrix}
    \tilde{p}_{N-i}(z_\alpha) \\ p_{a-1}(z_\alpha)
  \end{bmatrix} .
  \label{eq:inversedets_M>N}
\end{align}
To the best of our knowledge, the inverse characteristic polynomial formula for $M>N$ has not appeared before in the literature. It represents one of the main results of this paper.
We have thus seen how our analysis based on the Schur polynomial expansion provides a systematic method to calculate various averages of the characteristic polynomials, and their inverses.

\section{Generalizing to Non-Zero Source} \label{sec:withsource}

\subsection{Schur Polynomial Average with Source}

In order to calculate the characteristic polynomials with the source term, we first consider the (non-normalized) Schur polynomial average in the presence of an external matrix source,
\begin{align}
    \expval{s_\lambda(X)}_A = \int \dd{X} e^{-\Tr V(X) + i\Tr AX} s_\lambda(X)
    \, .
\end{align}
To evaluate this average, we use the Harish-Chandra--Itzykson--Zuber (HCIZ) formula~\cite{IZ}%

\footnote{
We remark
\begin{align}
    \frac{1}{\operatorname{Vol} U(N)}
    \int_{U(N)} \dd{U} e^{i\Tr X U Y U^{-1}}
    \ \xrightarrow{X \to \mathbbm{1}_N} \ \frac{1}{\operatorname{Vol} U(N)}
    \int_{U(N)} \dd{U} e^{i\Tr Y} = e^{i\Tr Y} .
\end{align}
This fixes the normalization of the integral.
}
\begin{align}
    \frac{1}{\operatorname{Vol} U(N)}
    \int_{U(N)} \dd{U} e^{i\Tr X U Y U^{-1}} = 
    \frac{\Gamma_2(N+1) (-i)^{N(N-1)/2}}{\Delta_N(X) \Delta_N(Y)} \det_{1 \le i, j \le N} \qty[ e^{i x_i y_i} ] ,
\end{align}
which yields
\begin{align}
    \expval{s_\lambda(X)}_A 
    & = \frac{C_N}{\Delta_N(A)} \frac{1}{N!} \int \prod_{i=1}^N \frac{\dd{x}_i}{2\pi} e^{-V(x_i)} \det_{1 \le i, j \le N} \qty[ e^{i\sa_i x_j} ] \det_{1 \le i, j \le N} \qty[ x_i^{\lambda_j + N - j} ]
    \nonumber \\
    & = \frac{C_N}{\Delta_N(A)}
    \det_{1 \le i, j \le N} \qty[ \braket{ e^{i\sa_i x} }{x^{\lambda_j + N - j}} ] .
    \label{eq:SchurPoly_av_A}
\end{align}
with $(\sa_{i})_{i = 1,\ldots,N}$ the eigenvalues of $A$ and the normalization constant
\begin{align}
    C_N = \operatorname{Vol} U(N) \times \Gamma_2(N+1) \times (-i)^{N(N-1)/2} = (2 \pi)^{N(N+1)/2} (-i)^{N(N-1)/2} .
    \label{eq:CN_const}
\end{align}
We use this result in the following to compute the characteristic polynomial, and its inverse, with non-zero source.

\subsection{Characteristic Polynomial Average with Source} \label{sec:detwithsource}

We first compute the expectation value of multiple determinant insertions, in the presence of source. 
In this Section, we reproduce the results of \cite{Kimura_2014}, in particular its Eq. (3.10). 

Applying the Schur polynomial expansion together with the formula~\eqref{eq:SchurPoly_av_A}, we obtain
\begin{align}
    \expval{\prod_{\alpha=1}^N \det(z_\alpha - X)}_A
    & = \frac{1}{\Delta_M(Z)} \sum_{\lambda \subseteq (M^N)} (-1)^{|\lambda|} \det_{1 \le \alpha,\beta \le M} \qty[z_\alpha^{N - \lambda_\beta^\text{T} + \beta - 1} ] \expval{s_\lambda(X)}_{A}
    \nonumber \\
    & = \frac{C_{N}}{\Delta_M(Z) \Delta_{N}(A)} \sum_{\lambda \subseteq (M^N)} (-1)^{|\lambda|} \det_{1 \le \alpha,\beta \le M} \qty[ z_\alpha^{N - \lambda_\beta^\text{T} + \beta - 1} ] \det_{1 \le i, j \le N} \left[ \braket{e^{i\sa_{i}x}}{x^{N+\lambda_{j}-j}}\right] .
\end{align}
We can again recognize this as the minor expansion on the first $M$ rows of the following $(N+M) \times (N+M)$ matrix,
\begin{align}
    &
    \sum_{\lambda \subseteq (M^N)} (-1)^{|\lambda|} \det_{1 \le \alpha,\beta \le M} \qty[ z_\alpha^{N - \lambda_\beta^\text{T} + \beta - 1} ] \det_{1 \le i, j \le N} \qty[ \braket{e^{i\sa_{i}x}}{x^{\lambda_j + N - j}} ]   
    \nonumber \\
    & = \det_{\substack{1 \le \alpha,\beta \le M \\ 1 \le i, j \le N}}
    \begin{bmatrix}
    z_\alpha^{N + M - \beta} & z_{\alpha}^{N - j} \\
    \braket{e^{i\sa_{i}x}}{x^{N + M - \beta}} & \braket{e^{i\sa_{i}x}}{x^{N - j}}
    \end{bmatrix}
    = \det_{\substack{1 \le \alpha,\beta \le M \\ 1 \le i, j \le N}}
    \begin{bmatrix}
    p_{N + M - \beta}(z_\alpha) & p_{N - j}(z_{\alpha}) \\
    \braket{e^{i\sa_{i}x}}{p_{N + M - \beta}} & \braket{e^{i\sa_{i}x}}{p_{N - j}}
    \end{bmatrix}
    \, .
\end{align}
where in going to the last equality, we again used the linearity of the inner-product and the invariance of the determinant under linear transformations on its rows to trade $x^{N+M-\beta} \rightarrow p_{N+M-\beta}(x)$.
We can thus write the final result as 
\begin{align}
    \expval{\prod_{\alpha=1}^N \det(z_\alpha - X)}_A
    & = \frac{C_{N}}{\Delta_M(Z) \Delta_{N}(A)} \det_{\substack{1 \le \alpha,\beta \le M \\ 1 \le i, j \le N}}
    \begin{bmatrix}
    p_{N + M - \beta}(z_\alpha) & p_{N - j}(z_{\alpha}) \\
    Q_{N+M-\beta}(\sa_{i}) & Q_{N - j}(\sa_{i})
    \end{bmatrix}
    \, 
\end{align}
where we borrowed the notation of \cite{Kimura_2014}
\begin{equation}
    Q_{k}(\sa)= \braket{e^{i\sa x}}{p_{k}}=\int \frac{\dd{x}}{2\pi} e^{-V(x)+ i\sa x}p_{k}(x) .
\end{equation}
This shows the formula presented in \eqref{eq:ChPoly_av_A}.

\subsection{Inverse Characteristic Polynomial Average with Source}\label{sec:invwithsource}

We now wish to generalize our calculation to the inverse characteristic polynomial with non-zero source
\begin{equation} \label{eq:withsource}
    \expval{\prod_{\alpha=1}^{M} \det(z_{\alpha}-X)^{-1}}_{A} = \int \dd{X} e^{- \Tr V(x) + i\Tr AX}\prod_{\alpha=1}^{M} \frac{1}{\det(z_{\alpha}-X)} .
\end{equation}
This computation is the main technical result of this paper.

With the expectation value of $s_{\lambda}(x)$ in the presence of the source $A$ as in~\eqref{eq:SchurPoly_av_A}, we use the Cauchy sum formula to expand the inverse determinants in terms of Schur polynomials. 
%
%
We begin with the case $M < N$. This means $\lambda_i = 0$ for $i > M$.
Then, we obtain a block matrix structure similar in spirit to the case without the source term \eqref{eq:SchurPoly_av_block},
\begin{align}
    \expval{s_\lambda(X)}_A 
    & = \frac{C_N}{\Delta_N(A)} \det_{\substack{1 \le \alpha, \beta \le M \\ M+1 \le a,b \le N}}
    \begin{bmatrix}
     \braket{e^{i\sa_\alpha x}}{x^{\lambda_\beta + N - \beta}} & \braket{e^{i\sa_\alpha x}}{p_{N - b}} \\
     \braket{e^{i\sa_a x}}{x^{\lambda_\beta + N - \beta}} & \braket{e^{i\sa_a x}}{p_{N - b}}
    \end{bmatrix} .
\end{align}
None of the blocks in the above matrix vanish. Hence, the determinant no longer factorizes into a determinant of the $M\times M$ and $(N-M)\times (N-M)$ diagonal blocks. This is the main complication relative to the case without source.

Using the same notation as in Section \ref{sec:invnosource}, we now want to perform the following sum over representations
\begin{align} \label{eq:hardsumreps}
   \frac{1}{M!} \sum_{\substack{0 \le r_{1},...,r_{M} \le \infty \\ r_{\alpha}\neq r_{\beta}}} \det_{1 \le \alpha, \beta \le M}\left[ z_{\alpha}^{M-N-r_{\beta}-1}\right] \det_{\substack{1 \le \alpha, \beta \le M \\ M+1 \le a,b \le N}}
   \begin{bmatrix}
     \braket{e^{i\sa_\alpha x}}{x^{\lambda_\beta + N - \beta}} & \braket{e^{i\sa_\alpha x}}{p_{N - b}} \\
     \braket{e^{i\sa_a x}}{x^{\lambda_\beta + N - \beta}} & \braket{e^{i\sa_a x}}{p_{N - b}}
    \end{bmatrix} .
\end{align}

After yet another application of the generalized discrete Andr\'{e}if--Heine identity and performing the geometric sum, we arrive at
\begin{align}
\expval{\prod_{\alpha=1}^{M} \det(z_{\alpha}-X)^{-1}}_{A} = & \frac{C_{N} }{\Delta_N(A) \Delta_M(Z)} \det_{\substack{1 \le \alpha, \beta \le M \\ M+1 \le a,b \le N}}
\begin{bmatrix}
  \braket{e^{i\sa_\alpha}}{ \frac{(x/z_{\beta})^{N-M}}{z_\beta - x} } & \braket{e^{i\sa_\alpha}}{ p_{N - b} } \\[.5em]
  \braket{e^{i\sa_a}}{ \frac{(x/z_{\beta})^{N-M}}{z_\beta - x} } & \braket{e^{i\sa_a}}{ p_{N - b} } 
\end{bmatrix} .
\end{align}
The final simplification arises by noting that we can rewrite
\begin{equation}
    z^{M-N}\left(\frac{x^{N-M}}{z-x}\right) = \frac{1}{z-x} +\sum_{s=0}^{N-M-1} \frac{x^{s}}{z^{s+1}} ,
\end{equation}
so that the first $M$ columns can be decomposed as 
\begin{align}
   z_\beta^{M-N} \braket{e^{i\sa_i x}}{\frac{x^{N-M}}{z_\beta - x}}
   & = \braket{e^{i\sa_i x}}{\frac{1}{z_\beta - x}}
   + \sum_{b=M+1}^{N} z_\beta^{-N+b-1} \braket{e^{i\sa_i x}}{x^{N-b}} .
\end{align}
The determinant is insensitive to this linear superposing of the other $N-M$ columns, so we can drop these other terms. Our final result therefore reads, as advertised: 
\begin{align}
 \expval{\prod_{\alpha=1}^{M} \det(z_{\alpha} -X)^{-1} }_A 
  = & \frac{C_{N}}{\Delta_N(A) \Delta_M(Z)} \det_{\substack{1 \le \alpha, \beta \le M \\ M+1 \leq a,b \leq N}}  \begin{bmatrix}
    \tilde{Q}(\sa_\alpha,z_\beta) & Q_{N-b}(\sa_\alpha) \\ \tilde{Q}(\sa_a,z_\beta) & Q_{N-b}(\sa_a)
    \end{bmatrix} 
\label{eq:inv_chpoly_M<N}
\end{align}
with the new function $\tilde{Q}(\sa,z)$ defined as in Eq.~\eqref{eq:tildeQ}.
The structure of this result agrees with a particular limit of a supermatrix model studied by one of the authors in \cite{2014} (see the $M=p=0$ limit of Eq. (3.3) in that paper). The normalization could not be computed there due to the use of a supermatrix generalization of the HCIZ angular integral.

\subsubsection{Zero Source Limit}\label{sec:zero_source_lim}

We can further check our formula by considering the limit of vanishing source, $A \rightarrow 0$. This limit is not completely straightforward because of Van-der-Monde determinant $\Delta_N(A)$ appearing in the denominator. We therefore need to expand the numerator to $\mc{O}(\sa_{i}^{N-1})$.
Begin by writing (the unusual labeling of $k$ in the Taylor expansion will be useful below) 
\begin{subequations}\label{eq:TaylorExp}
\begin{align} 
    \tilde{Q}(\sa_{i},z_{\beta}) = & \sum_{k=1}^{N} \frac{1}{(N-k)!} \tilde{Q}^{(N-k)}(0,z_{\beta}) \sa_{i}^{N-k}  + \mc{O}(\sa_{i}^{N}) , \\
    Q_{N-b}(\sa_{i}) = & \sum_{k=1}^{N} \frac{1}{(N-k)!} Q^{(N-k)}_{N-b}(0) \sa_{i}^{N-k} + \mc{O}(\sa_{i}^{N}) ,
\end{align}    
\end{subequations}
where 
\begin{subequations}
\begin{align}
    \tilde{Q}^{(k)}(0,z_{\beta}) = & \left.\frac{\partial^{k} \tilde{Q}(\sa_{i},z_{\beta})}{\partial \sa_{i} ^{k}}\right|_{\sa_{i}=0} = i^k \int \frac{\dd{x}}{2\pi} e^{-V(x)} \frac{x^{k}}{z_{\beta}-x} , \\
    Q^{(k)}_{N-b}(0) = & \left.\frac{\partial^{k} Q_{N-b}(\sa_{i})}{\partial \sa_{i} ^{k}} \right|_{\sa_{i}=0} = i^k \int \frac{\dd{x}}{2\pi} e^{-V(x)} x^{k} p_{N-b}(x) .
\end{align}
\end{subequations}
Interpreting the sum in the Taylor expansions of Eq.~\eqref{eq:TaylorExp} as matrix multiplication, and using $\det(AB)=\det(A)\det(B)$, we arrive at the following expansion for the determinant  
\begin{align}
&
\det_{\substack{1 \le \alpha, \beta \le M \\ M+1 \le a,b \le N}} \begin{bmatrix}
    \tilde{Q}(\sa_\alpha,z_\beta) & Q_{N-b}(\sa_\alpha) \\ \tilde{Q}(\sa_a,z_\beta) & Q_{N-b}(\sa_a)
    \end{bmatrix}  \nonumber \\
& = \det_{1\le i,j \le N} \left[ \frac{1}{(N-j)!} \sa_{i}^{N-j} \right] \det_{\substack{1 \le \alpha, \beta \le M \\ M+1 \le a,b \le N}}  \begin{bmatrix}
    \tilde{Q}^{(N-\alpha)}(0,z_\beta) & Q^{(N-\alpha)}_{N-b}(0) \\ \tilde{Q}^{(N-a)}(0,z_\beta) & Q^{(N-a)}_{N-b}(0)
    \end{bmatrix} + \mc{O}(\sa_{i}^{N}) .
\end{align}
The upper right block $Q_{N-b}^{(N-\alpha)}(0)$ vanishes by orthogonality if we specialize the monic polynomial $p_k(x) \to P_k(x)$, so that the second determinant factorizes into a determinant of each diagonal block. With the Van-der-Monde determinant written as $\Delta_N(A)=\det_{1 \le i, j \le N}(\sa_{i}^{N-j})$, and using the invariance of the determinant under linear transformation of the rows, we can simplify the above to 
\begin{align}
    \frac{ \tilde{\mc{Z}}_{N-M} }{\Gamma_2(N+1)(-i)^{N(N-1)/2}} \Delta_N(A) \det_{1 \le \alpha,\beta \le M} \left[ \int \frac{\dd{x}}{2\pi} e^{-V(x)} \frac{P_{N-\alpha}(x)}{z_{\beta}-x} \right]  + \mc{O}(\sa_{i}^{N})
\end{align}
where $\tilde{\mc{Z}}_{N-M} = \prod_{a=M+1}^{N}h_{N-a}$ comes from evaluating the determinant of lower right block. It is trivial now to take the limit $\sa_{i} \rightarrow 0$ and recuperate the result of Section \ref{sec:invnosource} (recalling the relative factor of $\mc{Z}_{N}$ used in the normalization there). 

\subsubsection{The Case with $M > N$}\label{sec:M>N}

So far, we have assumed the condition $M \le N$ in the computation. 
We now consider the opposite situation $M > N$, and calculate the inverse characteristic polynomial average.

In this case, we similarly apply the determinantal formula~\eqref{eq:block_Z_M>N}, and put to use the Schur polynomial average with source~\eqref{eq:SchurPoly_av_A}.
We then obtain
\begin{align}
    & \expval{\prod_{\alpha=1}^{M} \det(z_{\alpha}-X)^{-1}}_{A} 
    \nonumber \\ & 
    = \frac{C_N}{\Delta_M(Z) \Delta_N(A)} \sum_{0 \le \lambda_N \le \cdots \le \lambda_1 \le \infty} 
    \det_{\substack{i = 1,\ldots,N \\ \alpha = 1,\ldots,M \\ a = 1,\ldots,M-N}} \begin{bmatrix}
    z_\alpha^{-\lambda_i + i - (N+1)} \\ z_{\alpha}^{a-1}
  \end{bmatrix} 
  \det_{1 \le i, j \le N} \qty[ \braket{ e^{i\sa_i x} }{x^{\lambda_j + N - j}} ] .
\end{align}
Introducing $r_i = \lambda_i + N - i$ obeying $r_1 > r_2 > \cdots > r_N \ge 0$ as before, we can again use the generalized discrete Andr\'{e}if--Heine identity to write
\begin{align}
    &
    \sum_{0 \le \lambda_N \le \cdots \le \lambda_1 \le \infty} 
    \det_{\substack{i = 1,\ldots,N \\ \alpha = 1,\ldots,M \\ a = 1,\ldots,M-N}} \begin{bmatrix}
    z_\alpha^{-\lambda_i + i - (N+1)} \\ z_{\alpha}^{a-1}
  \end{bmatrix} 
  \det_{1 \le i, j \le N} \qty[ \braket{ e^{i\sa_i x} }{x^{\lambda_j + N - j}} ] 
  \nonumber \\ &
    = \det_{\substack{i = 1,\ldots,N \\ \alpha = 1,\ldots,M \\ a = 1,\ldots,M-N}}
  \begin{bmatrix}
    \displaystyle \sum_{r=0}^\infty z_\alpha^{-r-1} \braket{e^{i\sa_i x}}{x^r} \\ z_\alpha^{a-1}
  \end{bmatrix}
  = \det_{\substack{i = 1,\ldots,N \\ \alpha = 1,\ldots,M \\ a = 1,\ldots,M-N}}
  \begin{bmatrix}
    \tilde{Q}(\sa_i,z_\alpha) \\ p_{a-1}(z_\alpha)
  \end{bmatrix} .
\end{align}
To convert $z_\alpha^{a-1}$ to $p_{a-1}(z_\alpha)$ in the last equality, we have used the invariance of the determinant under linear transformation.
Hence, we arrive at the inverse characteristic polynomial formula for $M > N$,
\begin{align}
    \expval{\prod_{\alpha=1}^{M} \det(z_{\alpha}-X)^{-1}}_{A} 
    = \frac{C_N}{\Delta_M(Z) \Delta_N(A)} \det_{\substack{i = 1,\ldots,N \\ \alpha = 1,\ldots,M \\ a = 1,\ldots,M-N}}
  \begin{bmatrix}
    \tilde{Q}(\sa_i,z_\alpha) \\ p_{a-1}(z_\alpha)
  \end{bmatrix} .
  \label{eq:inv_chpoly_M>N}
\end{align}

\subsubsection*{Zero Source Limit}

The zero source limit $A \to 0$ of the formula~\eqref{eq:inv_chpoly_M>N} is similarly considered as in Section~\ref{sec:zero_source_lim}.
Applying the Taylor expansion for the function $\tilde{Q}(\sa_i,z_\alpha)$ with respect to the source \eqref{eq:TaylorExp}, we straightforwardly obtain
\begin{align}
    \lim_{A \to 0} \expval{\prod_{\alpha=1}^{M} \det(z_{\alpha}-X)^{-1}}_{A} 
    = \frac{\operatorname{Vol} U(N)}{\Delta_M(Z)}
    \det_{\substack{i = 1,\ldots,N \\ \alpha = 1,\ldots,M \\ a = 1,\ldots,M-N}}
  \begin{bmatrix}
    \tilde{p}_{N-i}(z_\alpha) \\ p_{a-1}(z_\alpha)
  \end{bmatrix} .
\end{align}
This reproduces the previous result \eqref{eq:inversedets_M>N}, again remembering the relative normalization constant $\mathcal{Z}_N$~\eqref{eq:Normalization_ZN}.

\subsubsection{Characteristic Polynomial/Source Duality}

The characteristic polynomial with the source shows an interesting duality under $Z \leftrightarrow A$~\cite{Brezin:2000CMP,Desrosiers:2008tp,Kimura_2014}.
We demonstrate this duality using the formula obtained in this paper.

Let us focus on the Gaussian case for simplicity, $V(x) = x^2/2$.
Then, $P_{k}$ and $Q_k$ are essentially equivalent to each other in this case.
Furthermore, we have
\begin{align}
    e^{\frac{1}{2}\sa^2} \tilde{Q}(\sa,z) 
    & = \int \frac{\dd{x}}{2\pi}  \frac{e^{-\frac{1}{2}(x-i\sa)^2}}{z - x}
    = \int \frac{\dd{x}}{2\pi}  \frac{e^{-\frac{1}{2} x^2}}{z - i\sa - x}
    = \int \frac{\dd{x}}{2\pi}  \frac{e^{-\frac{1}{2}(x+z)^2}}{-i\sa - x}
    = e^{-\frac{1}{2}z^2} \tilde{Q}(iz,-i\sa) . 
\end{align}
Comparing the two expressions for the inverse characteristic polynomial \eqref{eq:inv_chpoly_M<N} and \eqref{eq:inv_chpoly_M>N}, we explicitly see the duality $Z \leftrightarrow A$ in the Gaussian case. This duality had been derived for the Gaussian case via an integrating in-out-in-out procedure, rewriting the inverse determinants in terms of a bosonic integral (see section 4.2 of \cite{Kimura_2014}). 


\section{Discussion}\label{sec:discussion}

    \subsubsection*{Eigenvalue ``Fermi Sea'' Picture and the $M>N$ Regime for Inverse Determinant Correlators}
    
   The $M>N$ regime seems not to have been previously considered in the literature. One reason is perhaps that the condition $M\leq N$ or $M \geq N$ matters little for the better studied determinant insertions. There are multiple ways to see this. First, in the case of non-zero source, Eq.~\eqref{eq:ChPoly_av_A} is the determinant of an $(N+M) \times (N+M)$ matrix, with sub-blocks of size $M$ and $N$. Contrast this with the inverse characteristic polynomial result Eq.~\eqref{eq:inv_ch_poly_A}, which is the determinant of a $\operatorname{max}(M,N)$ sized matrix with a $\operatorname{min}(M,N)$ sized sub-block structure. Even in the case without source, $N$ is a \textit{lower} bound on the index of the orthogonal polynomials appearing in Eq.~\eqref{eq:justdets}, while it appears as an \textit{upper} bound in Eq.~\eqref{eq:justinversedets}. 
   
   In the ``Fermi Sea'' picture, the full probability distribution for the $n$ eigenvalues is nothing but the square of a Slater determinant ground state wavefunction for $n$ fermions in the potential $V(x)$. Eq.~\eqref{eq:justdets} may also be viewed as a Slater determinant wavefunction for fermions with position labels $(z_{\alpha})_{\alpha=1,\ldots,M}$. The role of the Van-der-Monde in the denominator is to cancel the $\Delta_M(Z)^2$ contained in the matrix measure $\dd{Z}$, against which we would normalize the square of the wavefunction (see the discussion around Eq.~(2.10) of \cite{Maldacena_2004}). One way to interpret the shift by $N$ in Eq.~\eqref{eq:justdets} is to say that determinant insertions correspond to particle excitations above the Fermi level. We can always populate higher energy levels, there is no cutoff. 
   
   Eq.~\eqref{eq:justinversedets} may then similarly be viewed as a Slater determinant wavefunction. From this perspective, inverse characteristic polynomials more closely resemble hole-excitations.\footnote{That said, this interpretation offers little as to the appearance of the Hilbert transform.} There is a natural cutoff, namely the depth of the Fermi sea. How then do we make sense of $M>N$? One speculative take on Eq.~\eqref{eq:inv_chpoly_M>N} is to say that we first make $N$ hole excitations. We then ``re-populate'' the Fermi sea with the remaining $M-N$ fermions, starting from the ground state. We identify these filled levels with the $(M-N)$-sized sub-block $\left[p_{a-1}(z_{\beta})\right]_{a=1,\ldots,M-N}$.

    \subsubsection*{String Theory Perspective on the Schur Expansion of Characteristic Polynomials}
    
    Matrix models provide one of the most concrete ways to study open/closed string duality. In particular, we can explicitly implement the idea of ``closing up holes'' on the open string worldsheet, replacing them with a superposition of closed string vertex operators. On the matrix side, this corresponds simply to rewriting the determinant operator as $\det(z-X) = \exp \Tr \log(z-X)$ and Taylor expanding for large $z$:
    \begin{equation} \label{eq:Branestoclosed}
        \prod_{\alpha=1}^{M} \det(z_{\alpha}-X) = \det_{M \times M}(Z) \exp \qty( -\sum_{k=1}^{\infty} t_{k} \Tr X^{k} )
    \end{equation}
    with 
    \begin{equation} \label{eq:Miwa}
        t_{k} = \frac{1}{k} \Tr_{M} (Z^{-k}) .
    \end{equation}
    The $t_{k}$'s are the so-called Miwa variables, and play the role of couplings for the closed string worldsheet ``vertex operators'' $\Tr X^{k}$. They play a crucial role in the Kontsevich duality \cite{Kontsevich:1992ti}. Eq.~\eqref{eq:Miwa} shows exactly how the insertion of D-branes can equivalently be re-expressed in terms of a new closed string background.
    
    
    One interesting take on the Schur expansion of the characteristic polynomial is as a decomposition of the brane insertions in terms of an orthogonal set of multi-closed-string ``operators''.\footnote{We thank Rajesh Gopakumar for discussion on this point.} Orthogonality is guaranteed by the character nature of the Schur polynomials under the $U(N)$ integral. We use the term ``multi-closed-string'' because the Schur polynomials, when written in terms of traces, contain products of traces. The role of the Miwa variable is played here by $s_{\lambda^{\vee}}(Z)$.\footnote{This Schur polynomial may be also interpreted as the Wilson loop contribution with respect to the background $U(M)$ gauge field~\cite{Ooguri:1999bv}.} This interpretation is inspired by \cite{BerensteinToy}, which discusses the ``Schur Polynomial Basis'' for the matrix quantum harmonic oscillator in terms of a closed string Fock basis.\footnote{Of course, we are dealing with a simple matrix integral here, and not a matrix quantum mechanics, so the notion of operator and state on the matrix side does not quite carry over (but they are sensible notions on the worldsheet).}

\section*{Acknowledgments}

EAM wishes to thank Bruno Balthazar, Rajesh Gopakumar, Jorrit Kruthoff and Phil Saad for useful discussions on the meaning of (inverse) characteristic polynomials, and their Schur expansion, in matrix models of string theory and JT gravity. EAM would also like to thank Pavel Wiegmann for a detailed discussion of orthogonal polynomials in RMT.
The work of TK was supported in part by ``Investissements d'Avenir'' program, Project ISITE-BFC (No. ANR-15-IDEX-0003), EIPHI Graduate School (No. ANR-17-EURE-0002), and Bourgogne-Franche-Comt\'e region.
EAM is supported by a Kadanoff fellowship at the University of Chicago, and would like to acknowledge the support of the Jones Endowment for Physics Research.

\bibliographystyle{utphys}
\bibliography{refs}

\end{document}